\documentclass[twocol]{ametsocV6.1}

\usepackage{lineno}
\usepackage{amsmath,amssymb}
\usepackage{booktabs}
\usepackage[greek, english]{babel}
\usepackage[
    colorlinks=true,
    linkcolor=red
]{hyperref}
\usepackage{academicons}
\definecolor{orcidlogocol}{HTML}{A6CE39}

\title{Baseline Climatology of Tornadoes and Waterspouts in the Philippine Archipelago}

\authors{Generich H. Capuli\href{https://orcid.org/0000-0003-1253-7043}{\includegraphics[scale=0.5]{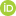}}\aff{a,b}\correspondingauthor{Generich H. Capuli, genecapuli@gmail.com / generich.capuli@pagasa.dost.gov.ph}
}
\affiliation{\aff{a}{Numerical Modelling Section, Research $\&$ Development and Training Division, Department of Science and Technology - Philippine Atmospheric, Geophysical, and Astronomical Services Administration, Brgy. Central, Quezon City 1100, Metro Manila, Philippines}\\
\aff{b}{Project Severe Weather Archive of the Philippines, Quezon City, Philippines}
}

\abstract{Tornadoes are documented on every inhabited continent, yet the environmental conditions used to anticipate them were developed and validated overwhelmingly on mid-latitude records, leaving tropical regimes largely unexamined. Here, we collected 1,186 combined tornado and waterspout reports in the Philippines spanning the early 1900s to 2025 and paired each with an ERA5 proximity sounding to construct the first baseline climatology and environmental characterization of these events in the archipelago. Activity concentrates in three corridors, mostly flanked by mountain ranges, one in each major island group, and is sharply phased within a May--October window that closely tracks the annual cycle of thunderstorm frequency, peaking in August and lagging into October for waterspouts alongside the tropical cyclone season. These events cluster in the mid-to-late afternoon, consistent with a daytime heating regime rather than nocturnal convection. The environments are thermodynamically permissive but kinematically shallow, with convective available potential energy (CAPE) exceeding typical Great Plains values, while the 0--6 km bulk wind difference (BWD; DLS) remains weak at $\sim$10--14 kt, departing from the mid-latitude covariance in which instability and shear increase together. Although streamwise vorticity is modest in magnitude, nearly all discriminating variance resides in the lowest kilometer, where 68\%--79\% of near-surface horizontal vorticity is streamwise, reaching $>$0.005 s$^{-1}$ and 96\% streamwiseness in the most organized profiles, and the hodographs retain low-level curvature. Consequently, composite indices remain low even as tornadoes and waterspouts recur. Mid-latitude-calibrated parameters can therefore misdiagnose tropical-maritime regimes, and recalibration toward shallow-layer, orientation-aware kinematic discriminators may recover the diagnostic signal.
}

\begin{document}

\maketitle

%


\section{INTRODUCTION}\label{sec1}

Severe convective storms exhibit pronounced regional variability throughout the world, governed by differences in regional climate and surface conditions \citep{Vengopal2016,Taszarek2021a} that partially determine which classes of severe weather occur in a given region. Among these, tornadoes represent one of the most extreme expressions of severe convective storms and rank among the most destructive meteorological phenomena, capable of inflicting substantial damage on buildings, infrastructure, and the landscape. Although documented on every inhabited continent \citep{GOLIGER1998,Maas2024}, tornadoes display marked regional contrasts in frequency, spatial distribution, and active season \citep{Niino1997,edwards2012,Farney2015,Antonescu2017,AndersonFrey2019,bai2020,Zhou2021}. The United States reports the most tornadoes, averaging more than 1,000 annually, whereas Europe records roughly a quarter of that total \citep{Grams2012,Antonescu2017}. Asia hosts several established hotspots \citep{Maas2024} such as in Japan \citep{Niino1997,kawazoe2023}, China \citep{Chen2018,Zhang2023,RZhang2025}, South Korea \citep{bai2020}, Indonesia \citep{Firdaus2025}, and the Bangladesh-India-Pakistan region \citep{bhan2016}. Still, their combined annual occurrence is approximately one-tenth of the U.S. figure \citep{Niino1997,fan2015,Zhou2021}.

Given that, it is widely accepted that tornadoes occur on every continent except Antarctica, and the Philippines is no exception; establishing a national tornado (including waterspout) climatology is therefore essential. Broadly, tornadoes are classified as supercellular (mesocyclonic) or non-supercellular (non-mesocyclonic). The former yield the most intense convective vortices and require the deep, persistent mesocyclone characteristic of supercell storms \citep{DAVIESJONES2015}. The latter encompass a range of vortices; such as landspouts and waterspouts, that arise instead from the intensification of pre-existing, shallow near-surface circulations \citep{Wakimoto1989}. Despite considerable progress, key aspects of tornado formation remain poorly understood, particularly where events are infrequent. Thus, characterizing the frequency, spatial distribution, seasonality, and environmental conditions of tornado occurrence not only elucidates the governing dynamics, but also underpins risk assessment and mitigation, strengthening community response, and helping to identify vulnerable populations \citep{Johnson2021}. However, Philippine tornado research remains in its infancy. Initial severe weather climatologies compiled by the community project Severe Weather Archive of the Philippines (SWAP), including for hailstorms, indicate that such phenomena are relatively common across Luzon, particularly the Greater Metro Manila Region (GMMR), Western Visayas, and Central Mindanao \citep{Capuli2024,Capuli2026}. Several significant events have been documented, predominantly in Luzon, including the EF1 tornado that struck the areas in Manila and Quezon City, Metro Manila, on 14 August 2016 \citep{capuli2026b}, and the localized tornado outbreak in Camarines Norte on 14 September 2025 \citep[13 September UTC;][]{Capuli2026c}, highlighting the occurrence of tornadoes across diverse geographical and environmental settings, particularly in areas where complex terrain and land-sea interactions may influence the development and organization of severe thunderstorms with entailed hazards. To date, however, no study has examined tornado and waterspout environments in the national context.

A substantial body of work has assessed the utility of sounding-derived parameters for distinguishing storm types and forecasting tornado occurrence, predominantly in the United States \citep{Rasmussen1998,Thompson2003,Nixon2022}. Reanalysis model-derived soundings and associated convective parameters are now routinely used to characterize tornado environments, which generally develop where high vertical wind shear, large convective instability, and abundant low-level moisture coincide. Wind shear is central to storm maintenance and intensification, reducing precipitation loading on the updraft and inducing a dynamic vertical perturbation pressure gradient force (VPPGF), so that organized, long-lived storms are favored under high shear \citep{Weisman1982,MARKOWSKI2009}. Hodograph shape further discriminates storm mode: long, straight hodographs are associated with splitting supercells, whereas strongly curved, sickle-shaped hodographs are widely regarded to favor cyclonic, right-moving supercell with tornado threat, especially when shear is strong near the surface \citep{Thompson2000,Coffer2020}.

The thermodynamic environment acts in conjunction with these kinematics in modulating severe storms and their hazards. Severe convection occurs across a broad spectrum of Convective Available Potential Energy (CAPE) and vertical wind shear, such that an abundance of one can offset a deficit of the other \citep{Johns1993}: strong tornadoes may require less shear when CAPE is high or less CAPE when shear is strong. Composite indices such as the Supercell Composite Parameter (SCP) and the Significant Tornado Parameter (STP), particularly when formulated with the Effective Inflow Layer, integrate kinematic and thermodynamic components and typically outperform individual parameters in discriminating storm type and tornadogenesis potential \citep{Rasmussen1998,Rasmussen2003,Thompson2003,Thomspon2012}. Such analyses have proven valuable in discriminating strong from weak tornadoes and resolving seasonal contrasts in tornadogenesis \citep{Rodriguez2021}, separating tornadic storms from other extreme wind events \citep{Shikhov2025}, revealing that operational forecasting parameters require regional recalibration \citep{Hanesiak2024,RZhang2025}, and evaluating projected changes in tornado environments under a warming climate \citep{kawazoe2023}.

Given these considerations, a baseline climatology is essential to support both operational forecasting and research into the environments conducive to tornadoes and waterspouts. In operational practice, convective instability and shear are often described qualitatively e.g., "marginal," "large," or "ample". Yet, in the Philippines, no comprehensive baseline exists to objectively anchor these quantifications, for either traditional or more recently developed parameters. This gap is especially consequential for tornadoes and waterspouts, whose environments in the Philippine setting can be different across various regions with well-established paradigm. Motivated by these limitations and conducted as Part III of Project SWAP, this study establishes, for the first time, a baseline climatology of sounding-derived parameters relevant to tornado- and waterspout-producing storms over the Philippines. Specifically, the study aims to:

\begin{enumerate}
    \item Characterize the environments supporting tornadoes and waterspouts,
    \item Assess the climatological occurrence and distribution of thermodynamic and kinematic parameters relevant to operational tornado and waterspout forecasting,
    \item Identify the climatologically large or extreme parameter distribution, and
    \item Examine the composite thermodynamic profiles and hodograph shapes associated with tornado- and waterspout-supportive environments.
\end{enumerate}

The paper is organized as follows; Section 2 outlines the materials and methods, including the reanalysis dataset and atmospheric profile parameters used in the study. Section 3 presents the thorough investigation and analysis. Finally, Section 4 provides the summary of findings and an extended discussion of the results.

\section{METHODOLOGY}\label{sec2}

\subsection{Tornado and Waterspout Event Data}

This study uses the 4th Data Release (DR4) of Project SWAP \citep{Capuli2024}, which compiles tornado and waterspout events from official and unofficial documentary sources to establish a baseline climatology. Official sources include DOST-PAGASA, the NDRRMC, DSWD-DROMIC, and local government units, while unofficial reports from news media, eyewitnesses, and social media contribute a substantial portion of the dataset, especially valuable for waterspouts and weak rural tornadoes that seldom enter official disaster records. Each report undergoes a three-stage verification process: (i) classification by source type, (ii) consistency checks on time and location, and (iii) metadata tagging (e.g., source link, platform, geo-coordinates, and visual documentation). Because both phenomena are visually distinctive and long-lived enough to be filmed, the proportion of visually confirmed events is high, and this imagery further supports discrimination of tornadoes from waterspouts. Additional pre-whitening is then applied: events identified as dust devils are removed, and reports with incomplete date or time information are excluded from the climatological environmental analysis, for which proximity sounding extraction requires a fully resolved timestamp, but retained for the temporal analysis where their metadata permit. Given the limited number of reports available in the DR4 dataset, additional tornado and waterspout accounts from non-official sources were identified and incorporated, including historical records dating as far back as the early 1900s. These supplementary records were included to extend the temporal coverage of the dataset and provide a more comprehensive representation of documented tornado and waterspout occurrences in the Philippines, particularly during periods when systematic observational records were scarce.

The imperfect nature of reporting affects all report-based datasets. Reported times, locations, and paths are not always accurate \citep{Verbout2006}, with report density being biased toward population. Hence, oversampling well-observed areas at the expense of sparsely settled uplands \citep{Groenemeijer2014,Antonescu2017}. For waterspouts, the bias is geometric as well as demographic since detection requires a shoreward line of sight or a nearby vessel, compressing the reported distribution toward coastlines and leaving offshore events largely absent \citep{Miglietta2018}. Daytime reporting bias is near-absolute for waterspouts, which are identified almost exclusively by sight \citep{Ashley2008}. Two errors are specific to the Philippine setting: (i) vernacular reports apply \textit{buhawi} and \textit{ipo-ipo} indiscriminately to tornadoes, waterspouts, and dust devils, so phenomenon assignment must be established from imagery during verification rather than inherited from the source, and (ii) damaging straight-line winds are frequently reported as tornadoes absent a formal survey \citep{TASZAREK2013}, a contamination the visual-evidence criterion suppresses but cannot eliminate. Intensity is not assigned, as light rural construction and offshore waterspouts provide damage indicators poorly suited to the Enhanced Fujita (EF) framework.

Despite these limitations, this citizen science approach offers a structured dataset that bridges a key observational gap in the Philippines. Similar frameworks have proven effective elsewhere; CoCoRaHS \citep{cifelli2005,reges2016}, mPING \citep{elmore2014mping,elmore2022mping}, and the European Severe Weather Database's media-derived tornado and waterspout climatology \citep{Groenemeijer2014,Antonescu2017}, demonstrating that verified, metadata-tagged community observations can meaningfully support severe weather studies in data-sparse regions \citep{tan2022}. As an example from a newspaper, a picture of a waterspout that occurred on August 1987 in Laguna de Bay and impacted parts of Las Pi\~{n}as area in Metro Manila is presented in Figure 1. In total, 1,186 severe weather events (tornadoes and waterspouts combined) were used to construct the climatology of tornadoes and waterspouts presented in Figure 2, including in its sub-figures. 

\begin{figure*}[!t]
\includegraphics[width=\textwidth]{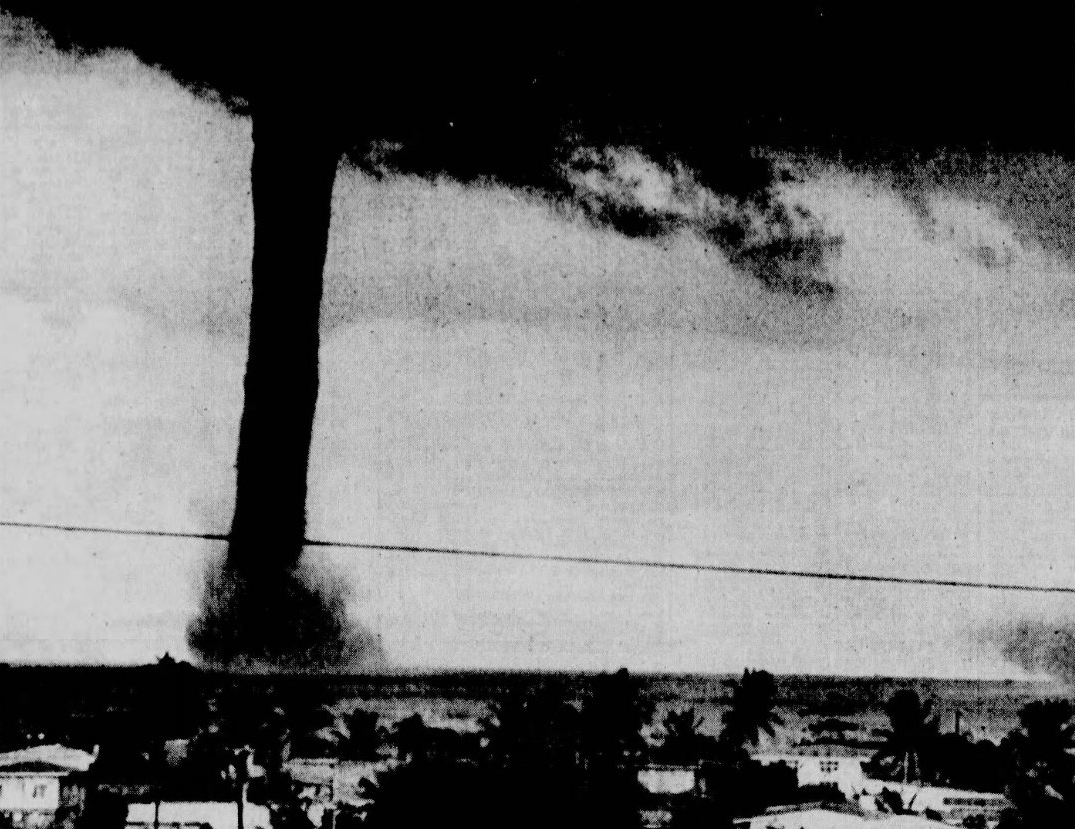}
\caption{Waterspout over Laguna de Bay on 6 August 1987 that impacted parts of Las Pi\~{n}as, Metro Manila. Picture courtesy of Reuters, published in The Buffalo News (issue of 8 August 1987).}
\label{fig_1}
\end{figure*}

\subsection{Environmental Reanalysis Dataset}

Reanalysis datasets are widely employed to diagnose atmospheric conditions conducive to severe weather such as tornadoes, damaging winds, and large hail. They are particularly useful for calculating instability indices and other parameters essential to severe thunderstorm forecasting \citep{Taszarek2020b}.

Environmental reanalysis data for this study were obtained from the fifth-generation European Centre for Medium-Range Weather Forecast (ECMWF) reanalysis (ERA5) accessed through the Climate Data Store \citep[CDS;][]{Hersbach2020}. ERA5 has demonstrated good performance in depicting vertical profiles of convective environments, especially across the United States and Europe \citep{Coffer2020,Taszarek2021a,Pilguj2022b}. However, some known biases persist, particularly in the boundary layer. These include discrepancies in low-level parcel characteristics and vertical shear parameters, especially near surface boundaries \citep{King2019}. Despite these limitations, ERA5 remains among the most reliable and accessible datasets for investigating severe convective environments, globally \citep{Coffer2020,Taszarek2021a,Taszarek2021b}. 

\subsection{Skew-T Hodograph}

To ensure comparable vertical resolutions upon angle-averaging, profiles used in composites were interpolated to a common 250 m. This spacing was chosen to ensure a balance between computational cost and resolution of the environmental features. Many obstacles to an advanced understanding of severe thunderstorm environments exist when considering ``close'' proximity soundings. Mainly, which time and space scales are most appropriate to represent the storm environment \citep{Brooks1994}. For example, \citet{Thompson2003} who used model soundings from Rapid Update Cycle-2 (RUC-2) analyses define a close proximity sounding as one taken within 40 km of the storm and within 3-hr of a standard sounding time (i.e., 00 and 12 UTC, including special soundings on 06 and 18 UTC). These authors took into account the influence of severe thunderstorms such as supercells in the low-level environmental profile up to 30 km, as demonstrated by \citet{weisman1998}. Here, we employed the approach of \citet{nixon2023} to define close proximity soundings, extracting vertical profiles from the ERA5 grid point closest to the severe weather event.

The parcel profiles were computed assuming a non-entraining, irreversible adiabatic process, following the recent formulation by \citet{Peters2022}. Compared to the pseudoadiabatic ascent where all condensate is assumed to fall out of an air parcel immediately \citep{Emanel1994}, the parcel calculation now accounts for the layer in which mixed-phase condensate is present just below the triple-point temperature. The most-unstable (MU) parcel profiles were used in this study since this can be used in both surface-based and elevated storm scenarios and detects the degree of low-level stability, and the lowest potential cloud base, respectively. This study also examines the relative humidity of the ambient air between 1--3 km and 1--6 km, a proxy for lower-tropospheric moisture above the cloud base where entrainment matters most \citep{Peters2019}, as a percentage from 0 to 100\%.

Kinematically, storm-relative quantities required assumptions on storm motion, as neither observed storm motions, supercell type (right- or left-moving), nor storm mode (supercell, multicell cluster, squall line) could be recovered from the documentary record. Estimated motions were therefore computed using the Bunkers ID method \citep[B2K;][]{Bunkers2000}, which yields both a right-moving (RM) and a left-moving (LM) vector relative to the non-pressure-weighted mean wind. Rather than default to the RM vector on hemispheric grounds alone, both branches were retained. Shear-based kinematics that are independent of storm motion (0--1 and 0--6 km AGL bulk wind difference/BWD) were computed directly from the hodograph.

Each event was further classified by the dominant ambient flow regime, following \citet{Capuli2026}, who delineated two severe weather seasons in the Philippines by the prevailing windflow in the atmospheric column: the easterly regime of the tropical trade winds (E) and the westerly regime of the southwesterly monsoon (W). Combining this two-way regime partition with the two Bunkers storm-motion branches yields four kinematic profiles per phenomenon: (i) easterly right-moving (E$\_$RM), (ii) easterly left-moving (E$\_$LM), (iii) westerly right-moving (W$\_$RM), and (iv) westerly left-moving (W$\_$LM), which form the organizational basis for the composite environments analysed below. All reanalysis-derived soundings associated with these events were analyzed using SounderPy by \citet{Gillett2025}.

\subsection{Thunder Hours}

A thunder hour is defined as an hour during which thunder can be heard \citep[e.g.,][]{jayaratne1998, bourscheidt2012}. Traditionally, thunder hours would be recorded by trained human observers. In this study, the thunder hour observations were derived using the observations gathered from the Earth Networks Global Lightning Network (ENGLN). The ENGLN dataset consists of lightning observations from nearly 2,000 Earth Networks Total Lightning Network (ENTLN) lightning sensors along with observations from over 70 World Wide Lightning Location Network (WWLLN) sensors. ENTLN utilizes ground-based broadband (1~Hz to 12~MHz) electric field change sensors to detect and locate both intra-cloud (IC) and cloud-to-ground (CG) lightning up to 1,500 km from the sensors. Meanwhile, WWLLN consists of ground-based very low-frequency (VLF) electric field sensors to locate primarily CG lightning on a global scale \citep{rodger2017, hutchins2012}.

Provided by \citet{digangi2022}, the thunder hour calculation was made using a $0.05^{\circ} \times 0.05^{\circ}$ latitude--longitude grid. For each grid point and for each hour, if at least two lightning pulses were located within 15 km of that point, the grid value was set to true (or 1), otherwise the grid value was false (or 0). The climatology of thunderstorm activity and its computation was made for all grid points and all UTC and local solar time (LST) hours from 1 January 2016 to 31 December 2025. The gridded data were then aggregated by calendar month. For each hour in a calendar month across all 10 years analyzed, the probability of thunder being observed at each grid cell was calculated. This is simply $N_{\mathrm{TH}}/N_{\mathrm{TOT}}$, where $N_{\mathrm{TH}}$ is the number of observations of thunder in a given hour of a given calendar month, and $N_{\mathrm{TOT}}$ is the total number of observations of the same hour of that calendar month. From these probabilities, the 95th percentile (P95) across all grid cells within the domain for each calendar-month hour was taken as a proxy for severe thunderstorm occurrence, isolating the upper tail of activity most plausibly associated with the deep convection from which tornadoes and/or waterspouts arise.

\subsection{Statistical Configuration}

The spatial climatology of tornadoes and waterspouts was constructed by kernel density estimation (KDE), applied independently to each phenomenon so that their geographies could be compared without one masking the other. For $n$ reports located at $\mathbf{x}_i$, the estimated density at any grid point $\mathbf{x}$ is;
\begin{equation}
  f(\mathbf{x}) = \frac{1}{n h^{2}} \sum_{i=1}^{n} K\!\left(\frac{d_i}{h}\right),
\end{equation}
where $d_i = \lvert \mathbf{x} - \mathbf{x}_i \rvert$ is the great-circle distance from the grid point to the $i$th report and $h$ is the bandwidth (search radius). A bivariate quartic (biweight) kernel was adopted in the form given by \citet{silverman1986};
\begin{equation}
  K(u) =
  \begin{cases}
    \dfrac{3}{\pi}\,\bigl(1 - u^{2}\bigr)^{2}, & \lvert u \rvert \le 1, \\[6pt]
    0, & \lvert u \rvert > 1.
  \end{cases}
\end{equation}
The quartic kernel weights each report smoothly to exactly zero at the search radius and is the standard used in comparable severe weather density climatologies \citep{Smith2012}. Estimation was carried out on a regular $0.05^{\circ}$ ($\sim$5 km) latitude--longitude grid covering the archipelago ($4.5$--$21.5^{\circ}$N, $116$--$127^{\circ}$E) using a fixed bandwidth of 50 km. Each field was then normalized by its own maximum and is therefore reported as a dimensionless relative density on $(0, 1)$, which permits direct comparison of the tornado and waterspout patterns despite their unequal sample sizes.

The seasonal cycle was described with a one-dimensional Gaussian kernel density estimate of event date, computed separately for the combined tornado-and-waterspout series and for the P95 thunder hour series, with the bandwidth set by Scott's rule ($h \propto n^{-1/5}$). A Gaussian kernel is used here, rather than the quartic kernel employed spatially, because the smoothed variable is a single continuous coordinate with no natural cut-off distance. The core of the season was delimited with the 70\% highest-density region \citep[HDR;][]{hyndman1996}, defined as the smallest set;
\begin{equation}
  R_{0.70} = \bigl\{\, t : f(t) \ge c_{0.70} \,\bigr\}
\end{equation}
whose enclosed probability equals $0.70$, with the threshold $c_{0.70}$ obtained by ordering the density values and finding the level at which the cumulative mass first reaches $0.70$.

Month of occurrence is a circular rather than a linear variable and seasonality was also treated with circular statistics \citep{mardia2000}. Each event was assigned an angle $\theta_i = 2\pi J_i / D$, where $J_i$ is the day of year and $D$ the number of days in that year. The mean resultant vector was computed from $C = \sum_i \cos\theta_i$ and $S = \sum_i \sin\theta_i$ as the mean resultant length;
\begin{equation}
  R = \frac{\sqrt{C^{2} + S^{2}}}{n},
  \qquad
  \theta_m = \operatorname{arctan2}(S, C),
\end{equation}
with circular mean $\theta_m$. $R$ measures the concentration of the distribution, approaching unity for a sharply phased season and zero for a distribution evenly spread around the year. The departure from uniformity was tested with the Rayleigh test, whose statistic $Z = n R^{2}$ has, under the null hypothesis of a uniform circular distribution, the approximate significance
\begin{equation}
  P \approx e^{-Z}
  \left[\, 1 + \frac{2Z - Z^{2}}{4n} \,\right].
\end{equation}

The correspondence between rotational severe weather and background thunderstorm activity was quantified on monthly aggregates in two complementary ways. First, ordinary least-squares linear and second-order polynomial (quadratic) models were fitted to monthly severe weather event counts as a function of monthly P95 thunder hours, and both the coefficient of determination $R^{2}$ and its adjusted form were reported. Second, the two annual cycles were compared as distributions rather than as a regression, using the overlap coefficient and the Bhattacharyya coefficient,
\begin{equation}
  \mathrm{OVL} = \sum_{m=1}^{12} \min(p_m, q_m),
  \qquad
  \mathrm{BC} = \sum_{m=1}^{12} \sqrt{p_m q_m},
\end{equation}
both evaluated over the twelve normalized monthly bins $p$ and $q$.

Differences between the parameter distributions of the four populations were tested with the two-sided Mann--Whitney U test \citep{Mann1947}, a rank-based test that makes no assumption of normality and is therefore suited to the skewed and zero-inflated distributions of the kinematic and composite parameters. Four planned contrasts were evaluated for each of the 20 parameters: easterly versus westerly tornadoes, easterly versus westerly waterspouts, and tornadoes versus waterspouts within each regime. The resulting 80 $p$-values were adjusted with the Holm step-down procedure \citep{Holm1979}. Effect size is expressed as Cliff's $\delta$ \citep{Cliff1993},
\begin{equation}
  \delta = \frac{2U_{1}}{n_{1} n_{2}} - 1 = P(X_{1} > X_{2}) - P(X_{1} < X_{2}),
\end{equation}
where $U_{1}$ is the Mann--Whitney statistic of the first group and $n_{1}$ and $n_{2}$ are the two sample sizes. $\delta$ ranges from $-1$ to $1$, with positive values indicating that the first group tends to take the higher values. Magnitudes were classified as negligible ($|\delta| < 0.147$), small ($< 0.33$), medium ($< 0.474$), or large ($\geq 0.474$) following \citet{Romano2006}. Signed kinematic and composite fields were compared as absolute values, consistent with their presentation in the violin plots.

\begin{figure*}[!t]
\includegraphics[width=\textwidth]{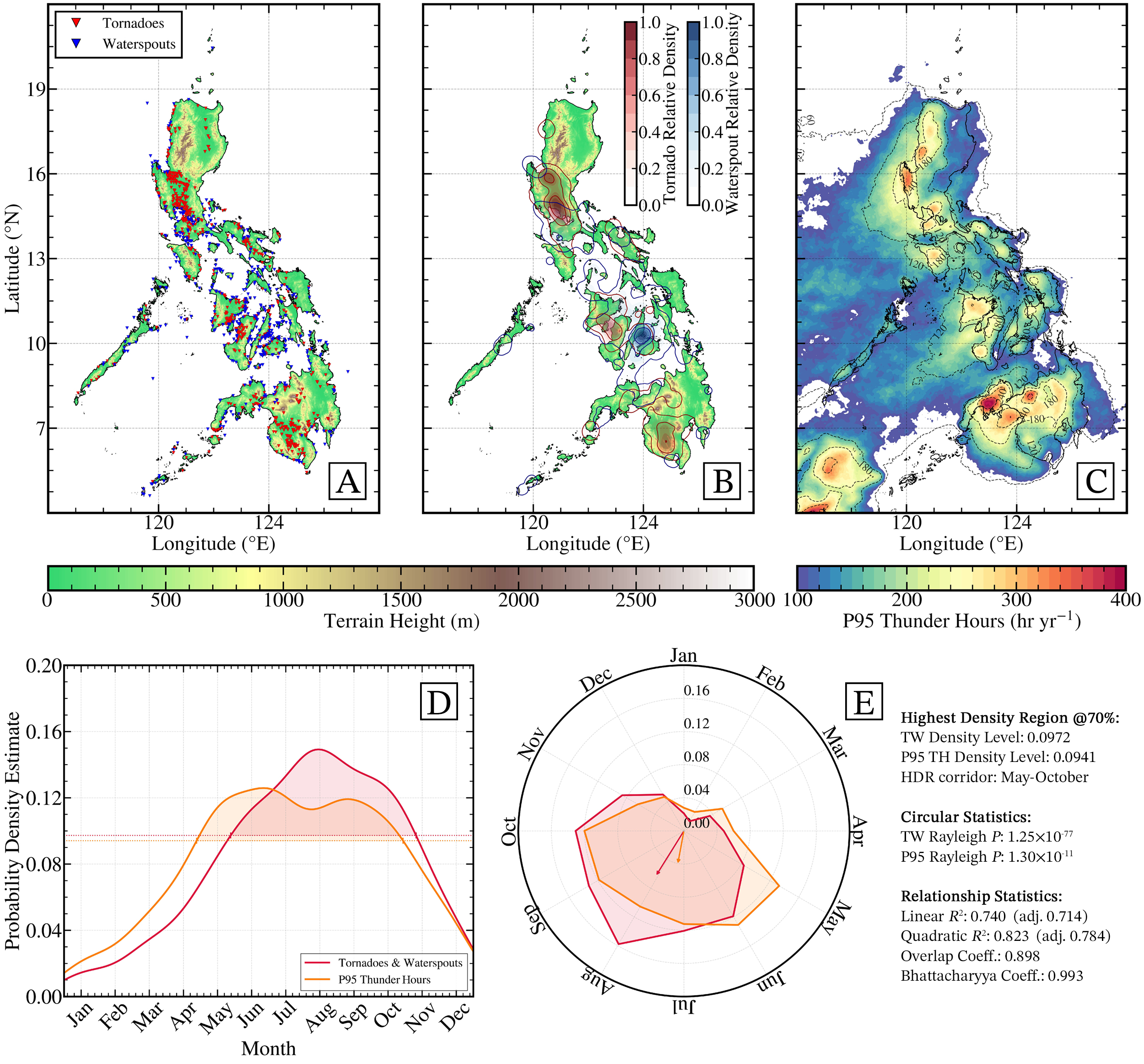}
\caption{(a) Spatial distribution of tornadoes (red triangle) and waterspouts (blue triangle) in the Philippines. (b) Climatological densities and hotspots of tornadoes (red line) and waterspouts (blue line) from 1900 to 2025. (c) 10-yr climatology (2016--2025) of P95 thunder hours (hr yr$^{-1}$) in the Philippines. Black dashed lines represent the mean thunder hours. Monthly probability density of Philippine tornado and waterspout events (crimson) and P95 thunder-hour days (orange), shown in: (d) linear and (e) circular form. Dotted lines in (d) denote the 70\% highest-density-region thresholds; arrows in (e) show the circular mean and resultant vector length. Summary statistics are included.}
\label{fig_2}
\end{figure*}

\section{RESULTS}\label{sec3}

\subsection{Spatio-Temporal Analysis of Tornadoes and Waterspouts}

Three hotspots emerge, one in each of the country's major island groups (Fig. 2a and 2b). The most pronounced lies in western Luzon, where tornado relative density exceeds 0.80 across two distinct regions extending from the lower Ilocos provinces; particularly Pangasinan, southward through the Mega Manila region to Batangas. Waterspout activity in Luzon is comparatively diffuse, concentrating around the southern of the two tornado maxima. A weaker but coherent signal in both phenomena extends across the Bicol Peninsula, from the Camarines provinces to Albay. The Luzon hotspot is flanked by the Zambales Mountain Range (ZMR) to the west and the Sierra Madre Mountain Range (SMMR) to the east. This terrain configuration is known to amplify rainfall from landfalling tropical cyclones through orographic forcing \citep{Lagmay2015,racoma2016}. Whether the same topography also pre-conditions the boundary layer for severe convection, on both TC and non-TC days remains untested, but the spatial coincidence is suggestive. In the Visayas, tornado activity centers on the western and central islands, with maxima exceeding 0.50 over Panay, Negros, and Cebu. A broader waterspout sector (relative density > 0.20) spans much of the same region, peaking above 0.80 over Cebu; the only location where the two distributions share a common maximum. The third hotspot occupies central Mindanao, peaking near 0.60 over South Cotabato and extending into North Cotabato, with a weaker extension into the northern part of the island. It too sits within a lowland corridor between two ranges: the Pantaron Mountain Range to the east and the Tiruray Highlands to the west. \citet{lagare2023} asserted that these complex topographic features in Mindanao promote the occurrence of mesoscale convective systems (MCS), thus may give an insight to the role of complex terrain along the said hotspot and for the increased tornadic activity in the area of interest.

Temporally, the Rayleigh test decisively rejects temporal uniformity for both series (tornadoes and waterspouts, $P$ = 1.25$\times$10$^{-77}$; P95 thunder hours, $P$ = 1.30$\times$10$^{-51}$), confirming a well-defined seasonal phase rather than a stochastic occurrence pattern. In Figure 2d, the 70\% highest-density region places the core of hazardous activity for both series within a May-October corridor, with comparable density levels (0.0972 and 0.0941, respectively), indicating that the seasonal envelope governing severe weather coincides with that governing the upper tail of thunderstorm frequency. This corridor aligns with the advection of the Asian summer monsoon to the Philippines i.e., Southwest Monsoon, including the peak season of tropical cyclone incursions, and the intervening monsoon breaks during which tropical easterlies prevail, each of which fosters severe convection through moisture supply, enhanced wind profiles and low-level convergence, and locally favorable instability \citep{Rasmussen1998,Thompson2003,Capuli2026}. 

The relationship statistics point to a direct link between overall thunderstorm activity and severe weather. Both the linear ($R^2$ = 0.740) and quadratic ($R^2$ = 0.823) fits explain a large share of the monthly variance, the modest quadratic improvement indicating a mild nonlinearity on an essentially monotonic increase. This supports the interpretation that the rise in tornado and waterspout activity is driven largely by the intensification of thunderstorm activity across the archipelago: as deep convection becomes more frequent during the monsoon and easterly-influenced months, the population of parent storms capable of producing these phenomena expands accordingly. The residual curvature reflects the additional requirement of favorable kinematic ingredients; vertical wind shear (e.g., low-, mid-, and deep-layer shear) and Storm-Relative Helicity (SRH), that thunder hour frequency alone does not capture \citep{Thompson2003,Thomspon2012,Coffer2020}. The distribution-similarity metrics reinforce this shared structure with overlap coefficient (0.898) and Bhattacharyya coefficient (0.993) both indicate near-coincident annual distributions, the latter approaching the theoretical maximum of unity.

Extended temporal analysis is performed as seen in Figure 3. The interannual record shows a steep rise in reported events after roughly 2019, culminating in 2024--2025 (Fig. 3a). A comparable incremental trend in tornado activity has been documented in US, European, and other national databases \citep{Verbout2006,Groenemeijer2014,LeonCruz2022,Sills2026}, where the apparent increase is generally attributed to the addition of more official reporting sources and heightened public interest in high-impact events rather than to a genuine climatological signal \citep{Antonescu2017}. The same pattern has recently been reported for tropical Indonesia, where a sharp post-2016 rise in tornado cases was linked to the widespread adoption of the internet and social media rather than an actual change in occurrence \citep{Firdaus2025}. The same interpretation applies to the Philippines as the observed rise is inferred to reflect enhanced social attention to these phenomena, the extensive use of social media networks, and increased access to internet services, rather than a true increase in occurrence. This reporting inhomogeneity, inherent to historical severe weather databases, warrants caution when interpreting the recent surge.

Severe weather activity is concentrated in the warm season, starting from March to April and extending from May through October with a pronounced August peak (Fig. 3b). During summer (JJA), the arrival of tropical waves and the onset of southwesterly winds \citep{dominguez2021} drive moisture transport from the surrounding oceans to the landmass; the resulting low-level moisture, instability, and favorable wind shear are the primary ingredients for the robust convective storms that produce severe weather hazards during this period \citep{Rasmussen1998,Thompson2003,DAVIESJONES2015,nixon2023}. This warm-season window is comparable to the tornado climatology of Mexico, where activity is concentrated from the end of spring through autumn, with a first peak in May and a second in July, likewise modulated by moisture advection from easterly waves and tropical cyclones \citep{LeonCruz2022,LEONCRUZ2025}. Waterspouts are phase-lagged relative to tornadoes, sustaining their maximum into October in coincidence with the tropical cyclone season \citep{teng2021,LeonCruz2022}. A comparable waterspout maximum has been reported in the Mediterranean, where activity peaks from summer into autumn under persistently warm sea-surface conditions \citep{Miglietta2018,Rodriguez2021}. This TC--waterspout association, although recognized in these regions and elsewhere \citep{weiss1987}, has not previously been documented for the Philippines and warrants further investigation. The concentration of activity within the monsoon season further echoes the tornado climatology of Indonesia, where the maximum coincides with the active phase of the Asia--Australia monsoon, albeit peaking in November under the boreal-winter monsoon rather than the boreal-summer window observed here \citep{Firdaus2025}. A minority of events in DJF and SON may correspond to cold-season tornadoes governed by distinct dynamics \citep{TASZAREK2013}. These regimes also partition by ambient flow: easterly-regime environments (Fig. 3c) dominate, with a broad June--October maximum consistent with prevailing easterly windflow and tropical easterly waves, whereas westerly-regime environments (westerlies and southwesterlies; Fig. 3d) are confined to July--September under the mature southwest monsoon and TC passages.

\begin{figure*}[!t]
\includegraphics[width=\textwidth]{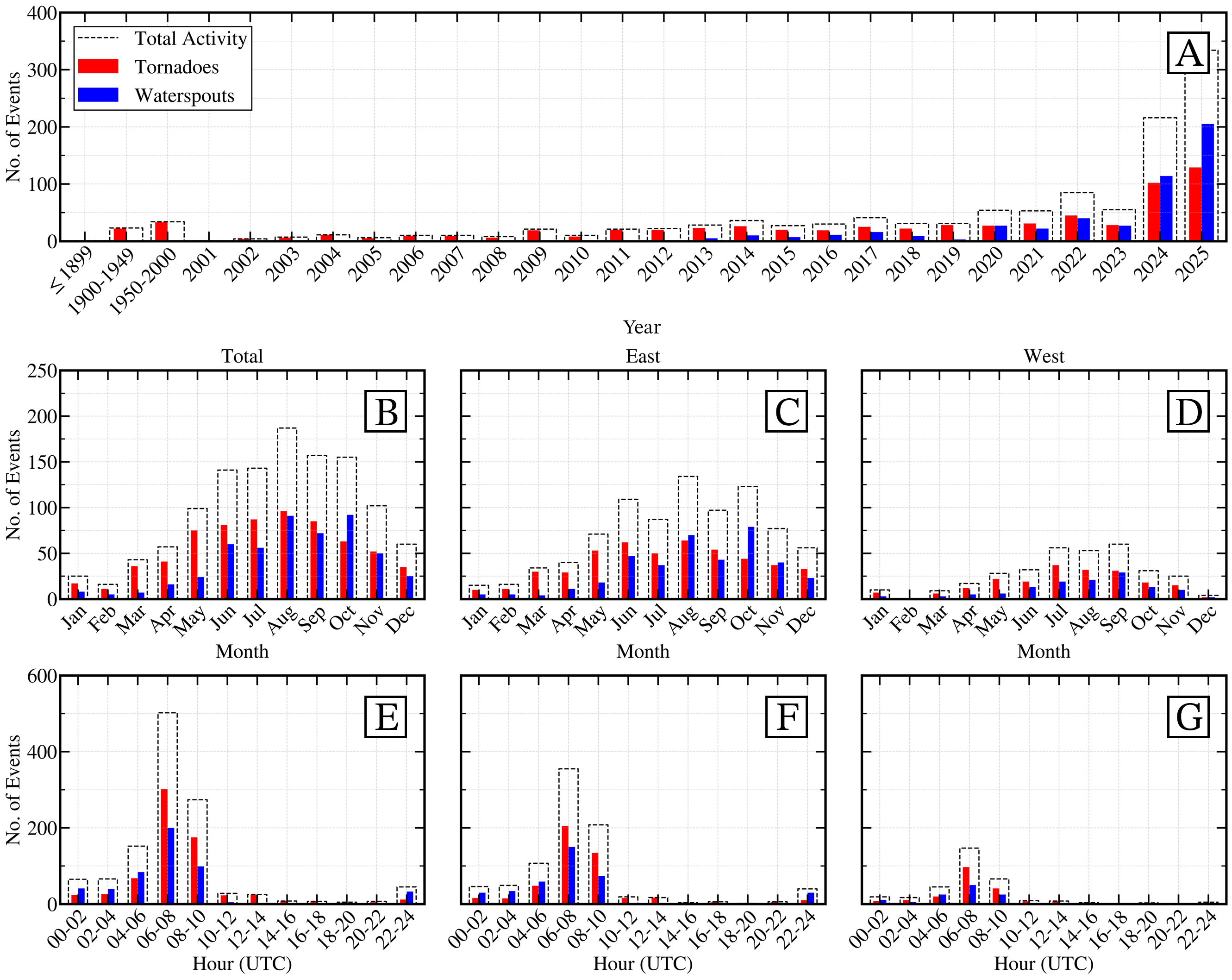}
\caption{(a) Yearly distribution of tornadoes (red) and waterspouts (blue) in the Philippines. (b) Monthly climatology of tornadoes and waterspouts, including as bisected into; (c) easterly wind regimes and (d) westerly wind regimes. (e) Diurnal climatology of tornadoes and waterspouts, including as bisected into: (f) easterly wind regimes and (g) westerly wind regimes. Black dashed lines represent the total activity.}
\label{fig_3}
\end{figure*}

The diurnal distribution (Fig. 3e) peaks between 14:00--15:59 local time (06:00--07:59 UTC), with a secondary maximum at 16:00--17:59 local time (08:00--09:59 UTC), reflecting daytime surface heating and air-mass thunderstorm development \citep{dai2001} that destabilize the boundary layer and drive convective persistence, updraft intensification, and tornadogenesis. This afternoon maximum is consistent with the diurnal cycle reported for Mexican tornadoes and Mediterranean tornadoes, whose activity shifts from a morning waterspout peak to an early-to-mid afternoon maximum for land-based severe weather events \citep{Miglietta2018,LEONCRUZ2025}, and with the tropical maritime climatology of Indonesia, where tornadoes concentrate between 13:00 and 17:00 local time under afternoon land--sea convective forcing \citep{Firdaus2025}. This signal is strongest under easterly-regime flow (Fig. 3f), which sustains both peaks and indicates prolonged late-afternoon convection, while westerly-regime environments (Fig. 3g) show a sharper, more singular peak consistent with their monsoon-forced convection. The scarcity of nocturnal events supports a pre-dominantly insolation-driven regime rather than one sustained by nocturnal low-level jets or elevated convection. Thus, matching several severe weather and rainfall climatologies \citep{Groenemeijer2014,Chen2018,Miglietta2018,banares2021,LeonCruz2022}.

\subsection{Climatological Distribution of Convective Parameters}

The thermodynamic environment is mainly defined by its instability. Figure 4 showcases the climatological distribution of thermodynamic and moisture parameters for tornadoes and waterspouts defined by their westerly and easterly-wind regimes. Tornadoes that developed in an easterly regime are accompanied by large instability, with a mean CAPE of 3547 J kg$^{-1}$ (IQR: 2592--4454 J kg$^{-1}$), exceeding westerly tornadoes (2690 J kg$^{-1}$; IQR: 1831--3598 J kg$^{-1}$) and both waterspout categories (IQR: 2616--2809 J kg$^{-1}$), and the easterly tornado distribution is also the broadest, its extensive IQR confirming that high buoyancy is not merely an average but a routine condition. These magnitudes place Philippine tornado environments at the high end of the global spectrum. \citet{Taszarek2020b} asserted that U.S. tornado environments are characterized by higher moisture, CAPE, wind shear, and mid-level lapse rates than their European counterparts, yet the mean easterly Philippine tornado exceeds typical Great Plains instability. This inverts the mid-latitude covariance in which, as \citet{Taszarek2020b} note, the seasonal increase in CAPE is accompanied by a decrease in shear and vice versa. Whereas, in the tropical maritime setting such as the Philippines, CAPE is greatest precisely where shear is least. High CAPE, however, is a weak discriminator here, echoing a broader pattern. Within Asia, \citet{Zhang2023} found that in China the thermodynamic parameters including CAPE, LCL, and cloud-level RH could not discriminate effectively between tornadic and non-tornadic supercells, noting that CAPE > 1000 J kg$^{-1}$ is already considered moderate to high after \citet{Rasmussen1998}.

The remaining thermodynamic fields reinforce a picture of deep, uniform tropical moisture. Mean LCLs sit firmly in the tornado-favorable range that mid-latitude climatologies identify. \citet{Rasmussen1998} found that more than half of significant-tornado soundings had LCLs below 800 m, and all four populations satisfy this, with easterly tornadoes at 603 m (IQR: 422--745 m) and the other three categories at or below $\sim$481 m and with tighter IQRs. The LCLs within environments of waterspouts are the lowest and least variable, consistent with maritime near-saturated inflow and with the shallow near-surface moist layer that \citet{LEONCRUZ2025} associates with landspout- and waterspout-type vortices forming from intensification of pre-existing low-level eddies rather than from mesocyclones. This maritime characteristic mirrors the Iberian and Balearic waterspout climatology of \citet{Rodriguez2021}, who likewise found that waterspouts occupy a distinct low-CAPE and form under environmental conditions markedly less restrictive than those required for tornadoes. Mean PWAT is high and narrowly clustered (55--59 mm, IQRs spanning only $\sim$8 mm), and mean column relative humidity is uniformly elevated, so the dry-air entrainment that disrupts low-level circulations in continental regimes is largely absent. Mean 0--3 km lapse rates are near moist-adiabatic and nearly identical across regimes (6.5--7.1 $^{\circ}$C km$^{-1}$) with IQRs under $\sim$1 $^{\circ}$C km$^{-1}$, offering no discriminating power. This near-uniformity is the sharpest contrast with both mid-latitude environments, where steep mid-tropospheric lapse rates from elevated mixed layers (EML) drive storm intensity and environmental variance \citep{Taszarek2020b}, and with the Luzon hail environments of \citet{Capuli2026}, in which steep mid-level lapse rates were a necessary co-ingredient, showing that lapse rate separates hazard types locally even as it fails to separate tornado from waterspout or easterly from westerly environments.

The kinematic environment is where the regime parts most clearly from the tornado-supporting environments. As shown in Figure 5, westerly-driven tornadoes average 12.5 kt (5.8--18.5 kt) in the 0--1 km layer, while easterly tornadoes average only 5.8 kt (2.8--7.4 kt), and the waterspout categories are weaker still. The wide, low-anchored IQRs indicate that a large fraction of even the westerly tornado population lies below the significant-tornado benchmark; roughly 10 m s$^{-1}$ (20 kt) of low-level shear has been proposed as a lower threshold for significant tornado events, with stronger shear associated with greater frequency of strong and violent tornadoes \citep{Rasmussen1998}. The contrast weakens with depth and reverses by 0--6 km, where the two tornado populations converge and the easterly regime is marginally deeper-sheared. \citet{Pilguj2022} report that violent European tornadoes were accompanied by 0--6 km wind shear $>$ 20 m s$^{-1}$, roughly three times the mean 0--6 km BWD in this study ($\sim$13 kt, or $\sim$7 m s$^{-1}$), extending across the region the shear deficit \citet{Taszarek2020b} documented in the mean.

\begin{figure*}[!t]
\includegraphics[width=\textwidth]{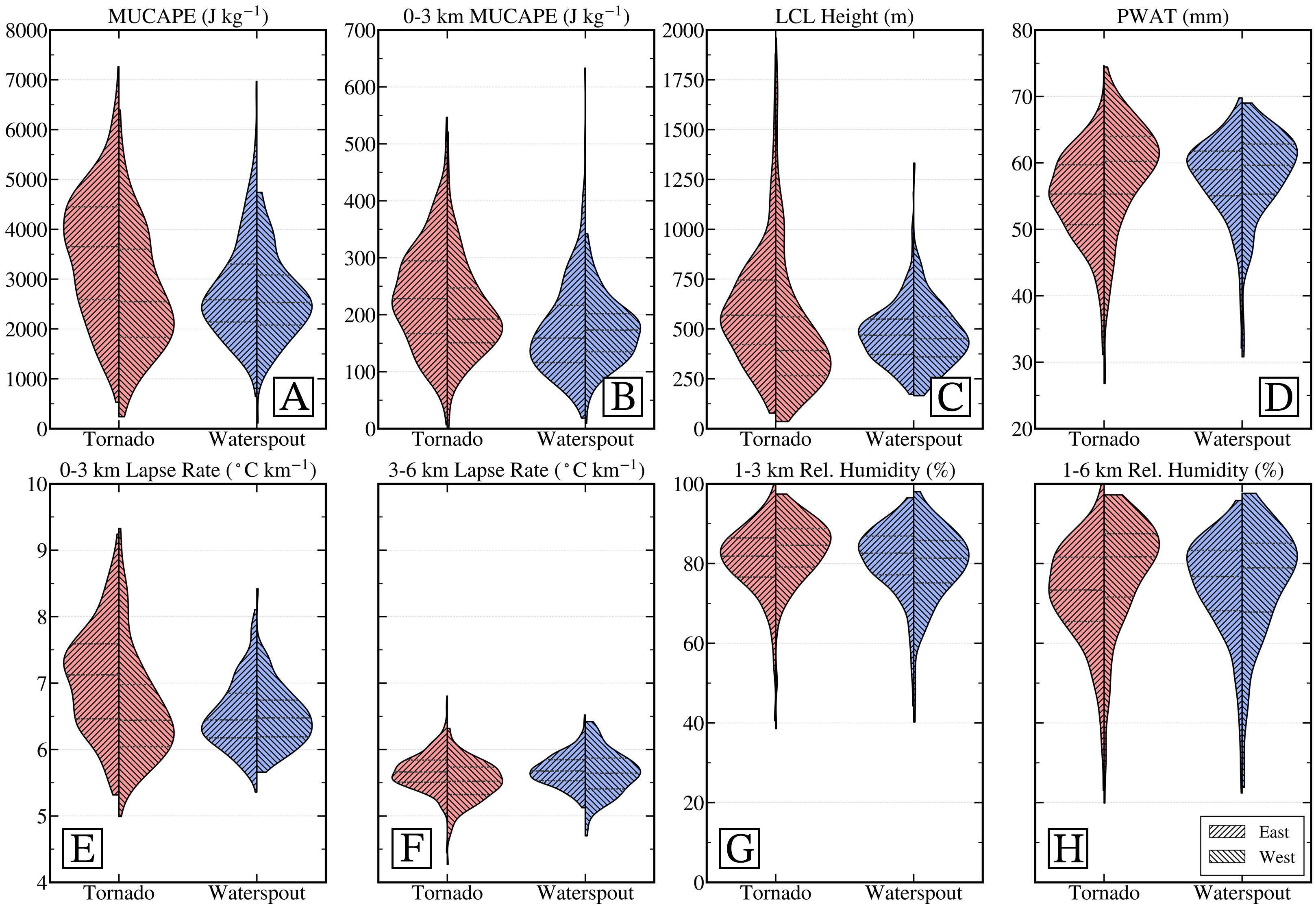}
\caption{Violin plots of tornadoes (red) and waterspouts (blue). Thermodynamic parameters are composed of: (a) MUCAPE (J kg$^{-1}$), (b) 0--3 km MUCAPE (J kg$^{-1}$), (c) LCL Height (m), (d) PWAT (mm), (e) 0--3 km Lapse Rate ($^{\circ}$C km$^{-1}$), (f) 3--6 km Lapse Rate ($^{\circ}$C km$^{-1}$), (g) 1--3 km Relative Humidity (\%), and (h) 1--6 km Relative Humidity (\%). Each is bisected based on easterly wind regimes (////) and westerly wind regimes (\texttt{\textbackslash\textbackslash\textbackslash\textbackslash}). 
}
\label{fig_4}
\end{figure*}

This can be also traced along the results of SRH in the archipelago. Westerly tornadoes lead at both depths (0--1 km SRH of 46.6 m$^{2}$ s$^{-2}$; 0--3 km of 69.3 m$^{2}$ s$^{-2}$), followed by westerly waterspouts, then the climatological easterly distributions. Read against operational thresholds derived from the U.S. record i.e., where 0--1 km SRH > 100 m$^{2}$ s$^{-2}$ and 0--3 km SRH $>$ 200 m$^{2}$ s$^{-2}$ suggest an increased tornado threat with supercells \citep{Thompson2003}, no climatological distribution approaches these values, even at the upper quartile. The interpretive weight placed on SRH is justified because \citet{Rasmussen1998} established that SRH discriminates between storm categories better than shear alone, implying that the streamwise component of horizontal vorticity dominates the production of rotating updrafts, as first shown by \citet{DaviesJones1984}. The layer over which SRH is most skillful is itself regionally contingent and shallow. \citet{Coffer2020} found that SRH integrated through the 0--500 m layer gave the greatest discrimination between significantly tornadic and non-tornadic supercells, and \citet{RZhang2025} independently found that in China 0--300 m SRH and bulk shear discriminated far better than deeper layers, refinements that likely apply to the similarly shallow-shear Philippine environments and merit direct testing. Waterspouts occupy the weak-helicity extreme of the distribution in both regimes (average easterly and westerly 0--1 km SRH of 15.3 and 26.2 m$^{2}$ s$^{-2}$; 0--3 km of 35.0 and 39.6 m$^{2}$ s$^{-2}$), closely matching the Iberian and Balearic waterspout climatological distribution as reported by \citet{Rodriguez2021}, where waterspout ranges of roughly 10--85 m$^{2}$ s$^{-2}$ for 0--1 km SRH and 35--110 m$^{2}$ s$^{-2}$ for 0--3 km SRH. The consistency of these weak-helicity ranges across two independent maritime regions reinforces that waterspout-producing environments are defined by their near-absence of low-level helicity rather than by any single limiting ingredient.

Streamwiseness of the horizontal vorticity behaves opposite to every other kinematic field as it is high rather than low. The mean fraction of horizontal vorticity that is streamwise is 68--79\% in the lowest 500 m, peaking in the westerly tornado subset with average of 78.6\% (IQR: 68.5--95.0\%), and even the lower quartiles remain well above 50\%. This implies that in a severe weather environment composed of weak low-level wind shear, it is the orientation of the vorticity, not its magnitude, that is favorable. The mechanism is well established: the tornadic supercells as simulated by \citet{Coffer2017} ingested pre-dominantly streamwise horizontal vorticity, promoting a strong low-level mesocyclone with enhanced dynamic lifting and stretching of surface vertical vorticity, whereas non-tornadic cases were dominated by crosswise vorticity in the lowest few hundred metres and failed to stretch surface vortices to tornadic intensity. The dimensional counterpart, mean streamwise vorticity, is set by the weak shear and is correspondingly small but not negligible: westerly tornadoes average 0.007 s$^{-1}$ (IQR: 0.002--0.011 s$^{-1}$) in the 0--500 m layer, roughly double the easterly value ($\sim$0.003 s$^{-1}$), with the westerly tornado upper quartile exceeding 0.010 s$^{-1}$. This streamwise vorticity > 0.005 s$^{-1}$ is broadly the level at which idealized and observational studies find low-level mesocyclones become robust enough for tornadogenesis, provided that low-level stretching is available to amplify surface vertical vorticity \citep{Coffer2017,Coffer2019,Finley2023,Fischer2024}. Vorticity stretching requires ample low-level buoyancy, and with sufficient 0--3 km CAPE across the climatological distributions (> 100 J kg$^{-1}$), that can supply the low-level updraft acceleration needed to convert well-aligned but weak streamwise vorticity into a vertical, eventually into a tornado as emphasized by \citet{Coffer2019} and \citet{Coffer2020}. Mean SRW, by contrast, is nearly uniform across populations (13--16 kt), indicating that inflow strength is not itself a distinguishing factor.

\begin{figure*}[!t]
\includegraphics[width=\textwidth]{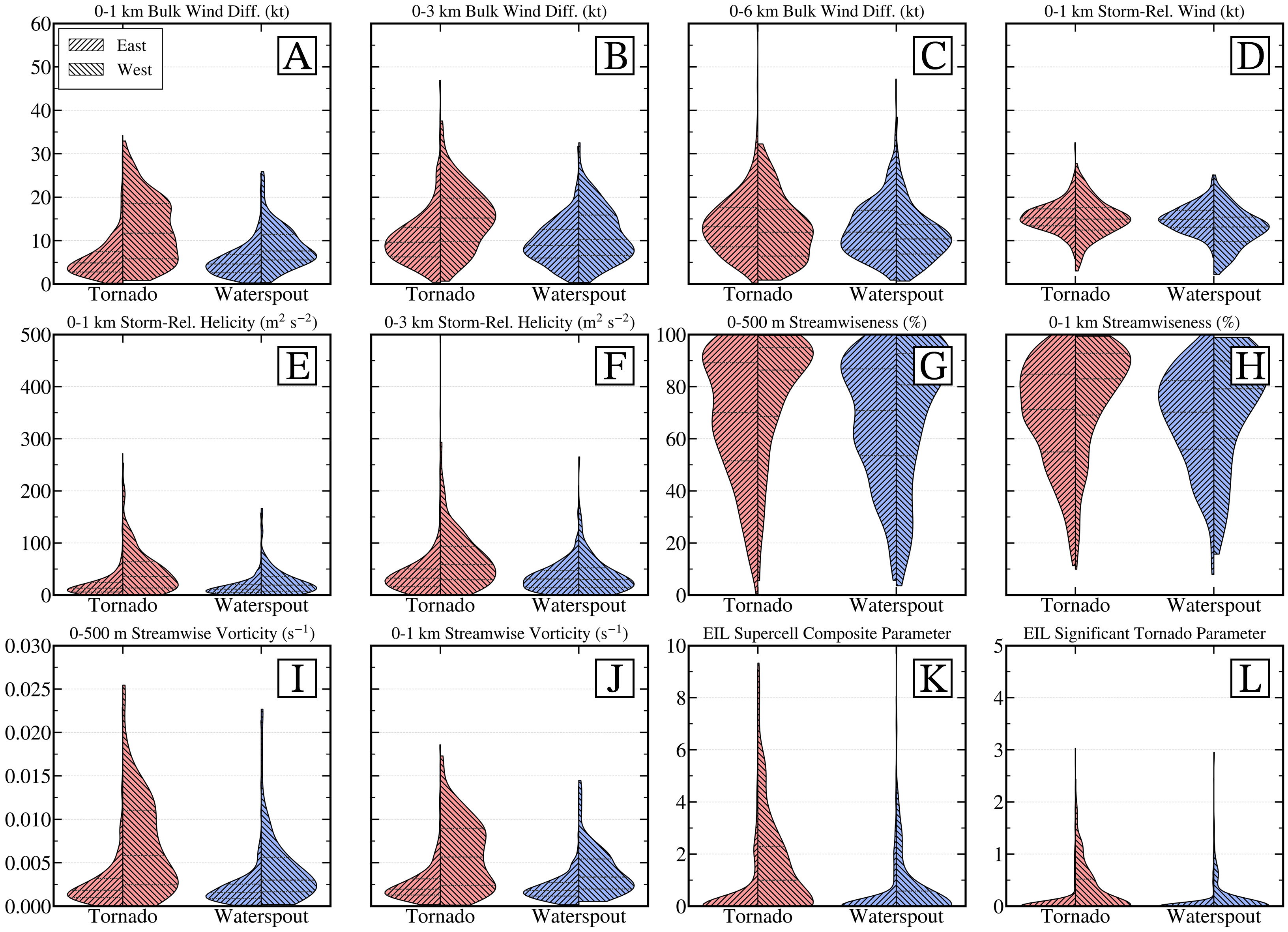}
\caption{As in Fig. 4, but for kinematic and composite parameters: (a) 0--1 km Bulk Shear (kt), (b) 0--3 km Bulk Shear (kt), (c) 0--6 km Bulk Shear (kt), (d) 0--1 km Storm-Relative Wind (kt), (e) 0--1 km Storm-Relative Helicity (m$^{2}$ s$^{-2}$), (f) 0--3 km Storm-Relative Helicity (m$^{2}$ s$^{-2}$), (g) 0--500 m Streamwiseness (\%), (h) 0--1 km Streamwiseness (\%), (i) 0--500 m Streamwise Vorticity (s$^{-1}$), (j) 0--1 km Streamwise Vorticity (s$^{-1}$), (k) Supercell Composite Parameter and (l) Significant Tornado Parameter both using Effective Inflow Layer and are dimensionless. Each is bisected based on easterly wind regimes (////) and westerly wind regimes (\texttt{\textbackslash\textbackslash\textbackslash\textbackslash}). 
}
\label{fig_5}
\end{figure*}

The composite indices are also modest, compared to other regions (Fig. 5k and 5l). Mean SCP is 0.49 and 1.55 for easterly and westerly tornadoes, respectively. Meanwhile, the mean STP is 0.08 and 0.32 respectively, which are values far below the canonical benchmarks in which, following \citet{Thomspon2012}, a majority of significant tornadoes (> (E)F2) are associated with STP > 1, while most non-tornadic supercells have STP < 1, and in which an SCP of 1 discriminates supercells from non-supercell storms. The zero-anchored lower quartiles show that most events register no supercell signal. The positive means are carried by the upper tail. Rather than dismiss these low values, we argue they retain diagnostic utility as regionally rescaled thresholds such as within this dataset the westerly tornado regime is uniquely distinguished by mean SCP approaching and exceeding unity and by a non-zero STP upper quartile. Locally, this means that SCP $\sim$ 1.0 and STP $\sim$ 0.5 can demarcate the transition into the most tornado-favorable environments the archipelago produces. This is supported by the case study of a localized tornado outbreak in Camarines Norte \citep{Capuli2026c}, in which SCP reached 2.6 and STP reached 0.6 near the time of the tornadoes, values described as modest but within the outer bounds of the tornado baseline climatology. Rescaling and recalibrating composite indices is not unique to the Philippines. \citet{Capuli2026} introduced a modified WMAXSHEAR as a useful discriminator for hail-producing storms. Furthermore, \citet{RZhang2025} showed quantitatively that the two original STP formulations achieved True Skill Score (TSS) of only 0.14 and 0.29 over China, and a China-calibrated formulation using shallow-layer SRH and shear raised the statistics to 0.51. Along the tropics, \citet{LEONCRUZ2025} reached the analogous conclusion for Mexico's predominantly non-supercell and supercell environments. While some environments did reach the thresholds, the composite parameters largely fail against U.S. benchmarks not because any single ingredient is absent, but because SCP and STP are multiplicative and reward the simultaneous presence of buoyancy and shear, and the defining regional feature is that these two ingredients rarely coincide. China \citep{Zhang2023}, Mexico \citep{LEONCRUZ2025}, and Luzon hail climatology \citep{Capuli2026}, converge on the inadequacy of U.S. Great Plains-calibrated indices in high-CAPE, low-shear regimes argues strongly for a recalibration built on shallow-layer, orientation-aware kinematic discriminators \citep{Rasmussen1998,Coffer2020,Zhang2023}.

\begin{figure*}[!t]
\includegraphics[width=\textwidth]{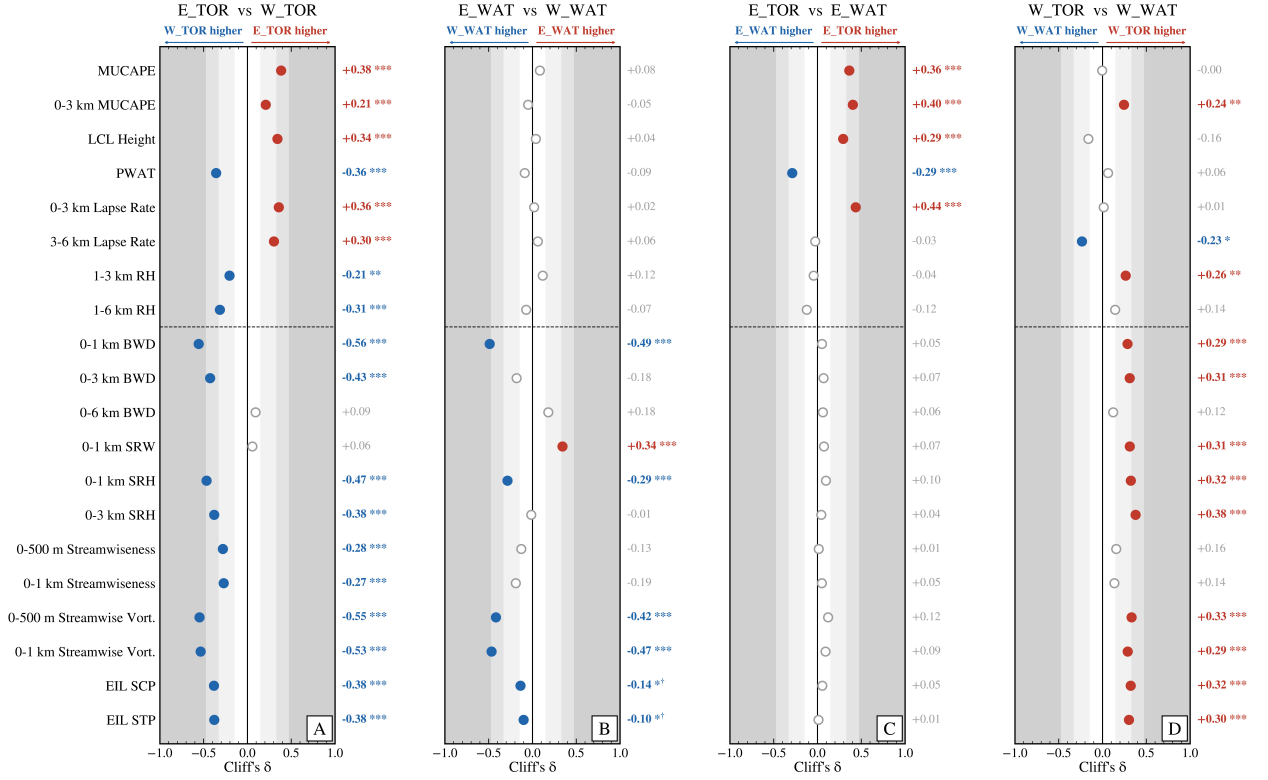}
\caption{Cliff's $\delta$ effect sizes from two-sided Mann--Whitney U tests comparing the environmental parameters of (A) easterly versus westerly tornadoes (E\_TOR, $n$ = 477; W\_TOR, $n$ = 201), (B) easterly versus westerly waterspouts (E\_WAT, $n$ = 382; W\_WAT, $n$ = 124), (C) easterly tornadoes versus easterly waterspouts, and (D) westerly tornadoes versus westerly waterspouts. Positive (negative) $\delta$ indicates that values in the first (second) group of each pair tend to be higher, as labelled by the arrows above each panel. Filled circles denote differences that remain significant after Holm correction across all 80 tests ($^{*}p < 0.05$, $^{**}p < 0.01$, $^{***}p < 0.001$); hollow grey circles are not significant. A dagger ($\dagger$) marks results whose significance changes when soundings are de-clustered to one daily median per group. Grey shading denotes effect magnitude following \citet{Romano2006}: $|\delta| < 0.147$ negligible (white), $< 0.33$ small, $< 0.474$ medium, and $\geq 0.474$ large (darkest). The dashed line separates thermodynamic (above) from kinematic and composite (below) parameters. Kinematic and composite parameters are compared as absolute values.}
\label{fig_mwu}
\end{figure*}

To test whether these contrasts are systematic rather than artefacts of the population means, the distributions in Figs.~4 and 5 were compared pairwise with the Mann--Whitney U test (Fig.~\ref{fig_mwu}). The test confirms that the deep layer carries no discriminating signal: the 0--6 km BWD differs negligibly in all four comparisons ($|\delta| \leq 0.18$, none significant), whereas the largest separation in the entire dataset lies in the lowest kilometer. Westerly tornado environments exceed their easterly counterparts in 0--1 km BWD ($\delta = -0.56$), 0--500 m streamwise vorticity ($\delta = -0.55$), and 0--1 km streamwise vorticity ($\delta = -0.53$), all large effects, followed by medium effects in 0--1 km SRH ($\delta = -0.47$), 0--3 km SRH, and both composite indices ($\delta = -0.38$). The thermodynamic environment is not, however, interchangeable between the regimes. Easterly tornadoes occur with systematically greater MUCAPE ($\delta = +0.38$), steeper 0--3 km lapse rates ($\delta = +0.36$), higher LCLs ($\delta = +0.34$), and lower PWAT ($\delta = -0.36$), medium effects that remain smaller than the leading kinematic ones. The two regimes therefore approach tornadogenesis from opposite ends of the buoyancy--shear trade-off described by \citet{Johns1993}: the easterly regime through buoyancy and low-level instability, the westerly regime through near-surface shear and streamwise vorticity. The fraction of soundings with non-zero composite indices expresses the same contrast, as 58\% of westerly tornado environments return SCP $>$ 0 and 47\% STP $>$ 0, against 21\% and 8\% of easterly tornado environments.

The tornado--waterspout contrast is itself regime-dependent. Within the easterly regime, the two phenomena are kinematically indistinguishable, with no kinematic or composite parameter differing significantly ($|\delta| \leq 0.12$), and are separated instead by thermodynamics: tornado environments have steeper 0--3 km lapse rates ($\delta = +0.44$), greater 0--3 km MUCAPE ($\delta = +0.40$) and MUCAPE ($\delta = +0.36$), and higher LCLs ($\delta = +0.29$), plausibly reflecting stronger low-level heating over land than over the adjacent sea. In the westerly regime the pattern reverses, as thermodynamic differences are mostly absent or small while tornadoes are distinguished by greater 0--3 km SRH ($\delta = +0.38$), 0--500 m streamwise vorticity ($\delta = +0.33$), SCP ($\delta = +0.32$), and 0--1 km SRH ($\delta = +0.32$). Waterspout environments, in turn, differ between the regimes only kinematically, through 0--1 km BWD ($\delta = -0.49$) and 0--1 km streamwise vorticity ($\delta = -0.47$), while every thermodynamic parameter is statistically indistinguishable ($|\delta| \leq 0.12$). Therefore, they refine the argument above, which is that wherever the westerly regime is involved, the discriminating signal resides in the low-level kinematics, whereas within the easterly regime it is buoyancy and low-level lapse rate, not shear, that separates tornado from waterspout environments.

\begin{figure*}[!t]
\includegraphics[width=\textwidth]{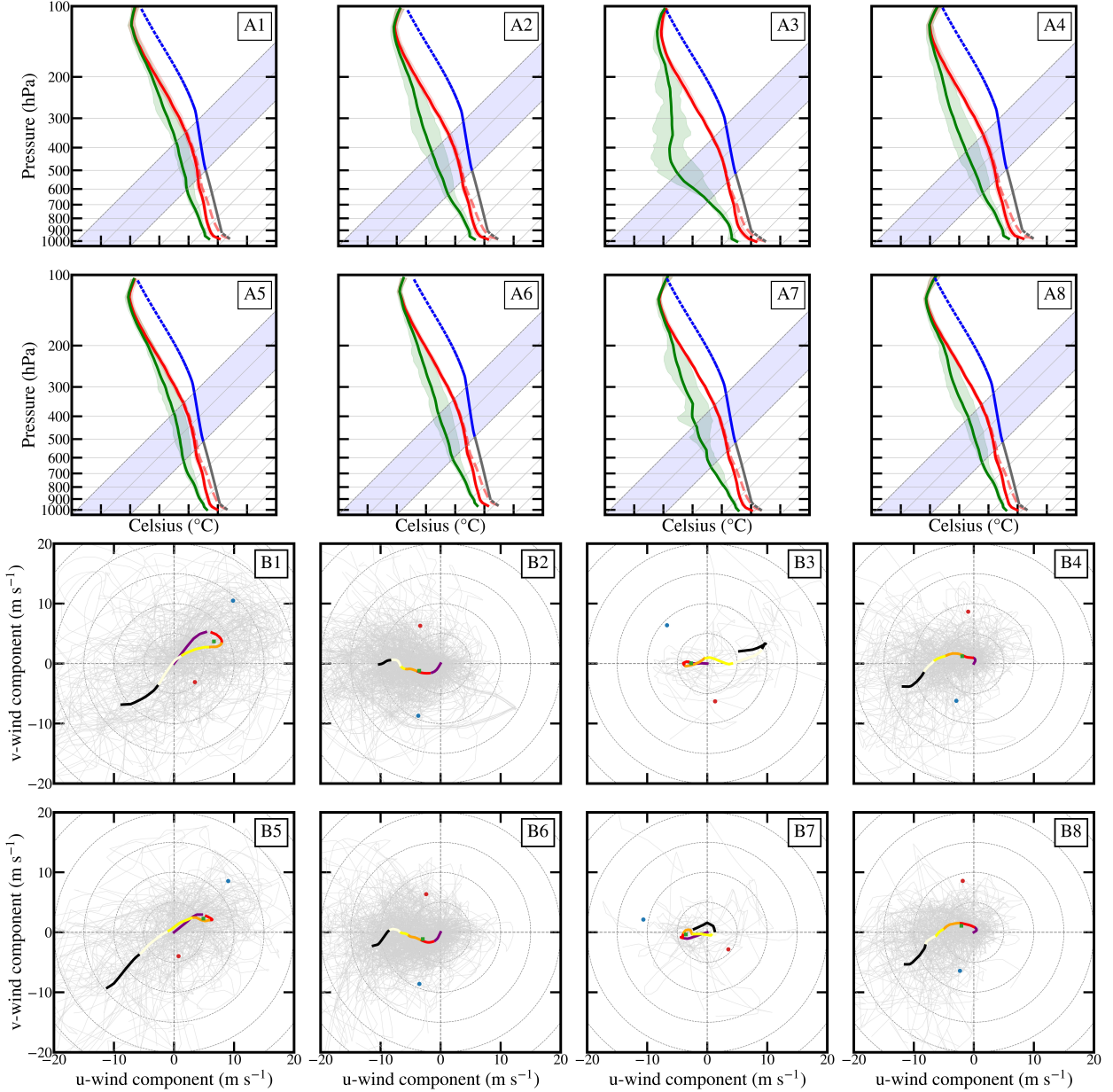}
\caption{Angle-averaged skew $T$--log$p$ ($\mathrm{A1}$--$\mathrm{A8}$) and hodograph ($\mathrm{B1}$--$\mathrm{B8}$) profiles for each dominant flow regime, classified by Bunkers storm-motion logic. Rows~1 and~3, tornadoes; Rows~2 and~4, waterspouts. Columns: (1)~$\mathrm{W_{RM}}$,
(2)~$\mathrm{E_{RM}}$, (3)~$\mathrm{W_{LM}}$, (4)~$\mathrm{E_{LM}}$, where W/E denote westerly/easterly flow and RM/LM the Bunkers right-/left-moving storm motion. 
}
\label{fig_6}
\end{figure*}

\begin{table*}[!t]
    \caption{Convective and kinematic parameters by domain and phenomenon. Statistical-based columns give the means of the violin-plot distributions across member soundings. Meanwhile, parcel-based columns (with storm-motion logic) are derived from parcel theory applied to the composite mean sounding, with storm-relative quantities referenced to the Bunkers right-moving (RM) or left-moving (LM) motion. Column labels combine flow regime (E/W), phenomenon (TOR/WAT), and storm motion (RM/LM).}
    \label{tab_1}
    \centering
    \resizebox{\textwidth}{!}{%
    \begin{tabular}{l cccc cccccccc}
    \hline\hline
     & \multicolumn{4}{c}{\textbf{Statistical-based}} & \multicolumn{8}{c}{\textbf{Parcel-based w/ Storm-Motion Logic}} \\
    \cmidrule(lr){2--5}\cmidrule(lr){6--13}
    \textbf{Parameter}
    & \textbf{E\_TOR} & \textbf{W\_TOR} & \textbf{E\_WAT} & \textbf{W\_WAT}
    & \textbf{W\_RM\_TOR} & \textbf{W\_LM\_TOR} & \textbf{E\_RM\_TOR} & \textbf{E\_LM\_TOR}
    & \textbf{W\_RM\_WAT} & \textbf{W\_LM\_WAT} & \textbf{E\_RM\_WAT} & \textbf{E\_LM\_WAT} \\
    \hline
    MUCAPE (J kg$^{-1}$)
    & 3546.570 & 2690.010 & 2808.977 & 2616.078
    & 2516.875 & 3002.981 & 3359.893 & 3522.973
    & 2363.914 & 2618.823 & 2857.469 & 2695.665 \\
    0--3 km MUCAPE (J kg$^{-1}$)
    & 235.606 & 205.921 & 174.786 & 172.213
    & 208.854 & 196.060 & 231.240 & 247.081
    & 169.269 & 166.566 & 185.207 & 170.857 \\
    LCL (m)
    & 602.596 & 480.616 & 476.530 & 471.190
    & 459.913 & 838.499 & 580.029 & 632.019
    & 451.282 & 588.374 & 460.398 & 489.650 \\
    PWAT (mm)
    & 54.874 & 58.790 & 57.883 & 58.359
    & 59.055 & 47.295 & 53.746 & 54.178
    & 58.089 & 53.086 & 57.023 & 57.429 \\
    0--3 km LR ($^{\circ}$C km$^{-1}$)
    & 7.072 & 6.617 & 6.528 & 6.498
    & 6.592 & 7.294 & 6.981 & 7.207
    & 6.473 & 6.733 & 6.550 & 6.556 \\
    3--6 km LR ($^{\circ}$C km$^{-1}$)
    & 5.678 & 5.514 & 5.685 & 5.656
    & 5.513 & 5.708 & 5.705 & 5.679
    & 5.646 & 5.730 & 5.688 & 5.707 \\
    1--3 km RH (\%)
    & 80.787 & 83.492 & 81.216 & 79.521
    & 83.393 & 75.895 & 79.996 & 80.231
    & 79.233 & 74.877 & 79.965 & 81.520 \\
    1--6 km RH (\%)
    & 72.369 & 78.237 & 74.481 & 75.327
    & 77.046 & 54.922 & 69.578 & 70.693
    & 73.671 & 64.661 & 71.552 & 73.309 \\
    0--1 km BWD (LLS; kt)
    & 5.786 & 12.511 & 5.150 & 8.735
    & 9.311 & 4.167 & 3.586 & 1.451
    & 6.425 & 3.851 & 2.750 & 1.555 \\
    0--3 km BWD (MLS; kt)
    & 10.432 & 15.307 & 9.467 & 11.480
    & 11.284 & 4.935 & 7.467 & 4.999
    & 7.045 & 3.676 & 6.435 & 6.266 \\
    0--6 km BWD (DLS; kt)
    & 13.717 & 12.715 & 13.016 & 11.082
    & 6.817 & 2.345 & 11.199 & 10.137
    & 1.723 & 0.867 & 10.378 & 10.927 \\
    0--1 km SRW (kt)
    & 15.506 & 15.140 & 15.074 & 12.648
    & 16.425 & 15.249 & 15.449 & 14.646
    & 16.259 & 14.966 & 14.956 & 14.460 \\
    0--1 km SRH (m$^{2}$ s$^{-2}$)
    & 20.129 & 46.579 & 15.288 & 26.240
    & 42.548 & 9.936 & 14.763 & 5.232
    & 26.171 & 3.712 & 9.584 & 4.954 \\
    0--3 km SRH (m$^{2}$ s$^{-2}$)
    & 40.667 & 69.287 & 35.020 & 39.556
    & 58.225 & 14.645 & 32.183 & 19.651
    & 26.364 & 8.410 & 27.716 & 25.219 \\
    0--500 m Streamwiseness (\%)
    & 67.798 & 78.577 & 67.865 & 70.758
    & 93.811 & 65.732 & 90.617 & 54.576
    & 96.883 & 6.050 & 72.532 & 23.721 \\
    0--1 km Streamwiseness (\%)
    & 69.212 & 77.550 & 67.906 & 72.573
    & 96.346 & 56.893 & 95.015 & 75.979
    & 83.954 & 26.585 & 82.091 & 57.680 \\
    0--500 m Streamwise Vorticity (s$^{-1}$)
    & 0.003 & 0.007 & 0.002 & 0.004
    & 0.0059 & 0.0017 & 0.0018 & 0.0004
    & 0.0041 & 0.0001 & 0.0009 & 0.0001 \\
    0--1 km Streamwise Vorticity (s$^{-1}$)
    & 0.003 & 0.006 & 0.002 & 0.004
    & 0.0050 & 0.0013 & 0.0018 & 0.0006
    & 0.0032 & 0.0004 & 0.0012 & 0.0006 \\
    EIL SCP
    & 0.492 & 1.545 & 0.315 & 0.626
    & 0 & 0 & 0 & 0
    & 0 & 0 & 0 & 0 \\
    EIL STP
    & 0.078 & 0.324 & 0.047 & 0.117
    & 0 & 0 & 0 & 0
    & 0 & 0 & 0 & 0 \\
    \hline
    \end{tabular}%
    }
\end{table*}

\subsection{Thermo-Kinematic Sounding Profiles}

The 4 $\times$ 4 composite (Fig. \ref{fig_6}) partitions the four severe weather patterns across the two ambient wind regimes and both Bunkers storm-motion branches, permitting a direct visual and parametric comparison against the established mid-latitude tornado-environments. Across all eight panels, the thermodynamic structure is remarkably uniform, whereas the kinematic structure carries essentially all of the variance. For a summary of measurements, this is presented in Table \ref{tab_1} along with the previous discussion on the climatological distribution for further comparisons.

The thermodynamic composites depict a consistently warm, moist, and highly unstable tropical troposphere. MUCAPE ranges from 2,000--3,500 J kg$^{-1}$, with the easterly-regime composites exhibiting the greatest instability; the lowest 3 km alone contributes 160--250 J kg$^{-1}$. These values substantially exceed the 1,000 J kg$^{-1}$ ``large-CAPE'' threshold of \citet{Rasmussen1998} and are generally greater than those reported for subtropical southern China by \citet{Zhang2023}, the closest global analogue considered here. The instability reflects warm, moisture-rich low levels coupled with relatively steep lapse rates, with 0--3 km lapse rates of 6.5--7.3 $^{\circ}$C km$^{-1}$, all exceeding the $\geq$ 6 $^{\circ}$C km$^{-1}$ threshold found in more than half of the central-U.S. cases of \citet{Zhang2023}. Although such lapse rates in the Great Plains are often linked to an elevated mixed layer, their presence here should not necessarily imply the same EML mechanism. 

Moisture is similarly pronounced with PWAT of 47--59 mm and LCLs generally between 400--600 m. The slightly higher LCLs in the left-moving tornado and waterspout composites coincide with comparatively drier mid-levels. Nevertheless, the cloud-bearing layer remains sufficiently moist, with RH of 74--83\% at 1--3 km and 54--77\% at 1--6 km. The presence of ambient RH reduces the potential for substantial buoyancy dilution through entrainment, which can otherwise weaken convective updrafts by incorporating environmental dry air into the rising parcel \citep{Peters2023b}. The combination of high low-level and modest mid-level moisture, including low LCLs, favors sustained buoyant updrafts rather than strongly entrainment-limited convection.

Collectively, these profiles occupy the low-LCL, moist-inflow, weakly inhibited end of the tornado-environment spectrum described by \citet{Rasmussen1998} and \citet{Thomspon2012}, closely resembling the thermodynamic environment of the Camarines Norte easterly wave outbreak documented in this project \citep{Capuli2026c}. Unlike the ``loaded-gun'' environments common to the U.S. Great Plains, CIN is nearly absent across these tropical profiles. Consequently, thermodynamic sufficiency appears to be a background characteristic rather than a limiting ingredient. This is indicative that convection can initiate readily, making the subsequent tornado or waterspout outcome more dependent on the kinematic and mesoscale environment than on the availability of buoyancy itself.

In conjunction, the angle-averaged hodographs reveal that the kinematic conditions are weak only against the mid-latitude supercell counterparts, and reading them solely through that lens obscures their most important feature. DLS is modest, only slightly exceeding 10 kt (10.1--11.2 kt) in all easterly-regime composites and near-negligible ($<$ 7 kt) in the westerly cases, both far beneath the DLS range reported by \citet{Zhang2023} across China and the two U.S. regions. But the deep-layer metric is the wrong diagnostic here. What the hodograph profiles retain, and retain systematically, is genuine low-level curvature and shear. The hodographs are not the long, straight profiles that \citet{Nixon2022} associate with splitting, hail-favoring storms. Instead, they are accompanied by short, curved wind profiles whose bending and vector length are concentrated in the lowest one to three kilometers. In the most strongly organised environment; W\_RM\_TOR, LLS reaches $\sim$9.3 kt and MLS $\sim$11.3 kt, the largest values among the composites, atop a hodograph whose curvature mirrors, in form, if not in magnitude, the clockwise turning that both \citet{weisman1986} and \citet{Thompson2000} identify as the kinematic signature of the cyclonic, right-moving supercell. This shows that the tornadic character of these environments is encoded in the shape and depth of the low-level hodograph rather than in the length of the DLS vector, including its values.

Crucially, the vorticity these curved profiles contain is not merely present but efficiently oriented for updraft rotation. The 0--1 km streamwiseness exceeds 70\% in five of the eight composite profiles and surpasses 90\% in the more organised right-moving tornado cases (96.3\% for W\_RM\_TOR and 95.0\% for E\_RM\_TOR), remaining high (82\%--84\%) in their right-moving waterspout counterparts, while the 0--1 km streamwise vorticity peaks at $\sim$5.0 $\times$ 10$^{-3}$ s$^{-1}$ in the W\_RM\_TOR profile and $\sim$3.2 $\times$ 10$^{-3}$ s$^{-1}$ in its waterspout counterpart, falling by an order of magnitude toward the disorganised easterly and left-moving cases ($\sim$4--6 $\times$ 10$^{-4}$ s$^{-1}$). \citet{DaviesJones1984} established that it is streamwise vorticity, tilted and stretched by the updraft, that generates updraft rotation, and the high streamwiseness here indicates that what little low-level vorticity these tropical environments possess is almost fully available for that conversion. 

This shifts the diagnostic emphasis decisively into the 0--1 and 0--3 km layers, where the population variance is in fact organized. The W\_RM wind profiles exhibit the largest SRH (0--1 km $\sim$ 42.5 m$^{2}$ s$^{-2}$ for tornadoes, $\sim$ 26.2 m$^{2}$ s$^{-2}$ for waterspouts; 0--3 km SRH $\sim$ 58.2 m$^{2}$ s$^{-2}$ and $\sim$ 26.4 m$^{2}$ s$^{-2}$), while the easterly and left-moving composites collapse toward single-digit or near-zero helicity. This westerly-over-easterly and RM-over-LM ordering is the kinematic expression of the ambient-flow partition that structures the climatology, localising the rotational potential in the monsoon-westerly regime rather than the tropical-easterly one. A LLS magnitude near 10 kt appears to mark a practical floor for the organised and potential for a tornadic environment: only the westerly right-moving tornado environment clears it in the 0--1 km layer, and it is precisely this composite that carries the highest streamwiseness, streamwise vorticity, and helicity. This is consistent with the near-ground focus of \citet{Coffer2020}, who showed that helicity integrated over a shallow surface-based layer carries the operative tornado-forecast skill. The waterspout composites reinforce the point: their distinguishing shear and streamwise vorticity are the shallowest of all, and their environments, like the weak-tornado cases of \citet{Groenemeijer2014} and \citet{Taszarek2020b} in Europe, maximize where low-level lapse rates are steep and ample low-level CAPE despite weak DLS.

\section{SUMMARY AND CONCLUSION}\label{sec4}

The climatology of tornadoes and waterspouts, and their respective convective environment assembled here establishes that Philippine tornadoes and waterspouts occupy a corner of the global severe-storm parameter space that is systematically under-represented in the literature on which operational tornado diagnostics were built. The defining feature is not any single extreme value but a structural decoupling: thermodynamic conditions are uniformly permissive across every phenomenon, flow regime, and storm-motions, while kinematic conditions; weak in the deep layer but organized and orientation-favorable near the surface, carry essentially all of the discriminating variance. This inverts the covariance that underpins mid-latitude severe-storm forecasting, where instability and shear tend to vary together seasonally and where the strongest events emerge from their coincidence. In the Philippine setting, buoyancy is greatest precisely where shear is least, so the multiplicative logic embedded in the SCP and STP; which reward the simultaneous presence of both ingredients, collapses toward zero even as genuine tornadoes and waterspouts continue to occur. This is the same failure mode that \citet{Zhang2023,RZhang2025} diagnose for China and that \citet{LEONCRUZ2025} reports for Mexico, and its recurrence across three independent tropical and subtropical regions indicates a general limitation of U.S. Great Plains-calibrated indices rather than a regional peculiarity.

The apparent paradox of organized rotational storms occurring despite DLS an order of magnitude below conventional supercell thresholds is resolved by the shallow layer that deep-layer metrics largely obscure. Although BWD becomes negligible above the lowest few kilometers, the hodographs retain appreciable low-level curvature, with mean 0--500 m streamwiseness of 68--79\% and values exceeding 90\% in the most organized right-moving composites, making much of the available vorticity favorably oriented for conversion into updraft rotation. This structure is consistent with the idealized simulations of \citet{Coffer2017}, which showed that tornadic storms preferentially ingest near-surface streamwise vorticity, and with the emphasis on shallow-layer kinematics in \citet{Coffer2020} and \citet{Zhang2023}. In the westerly tornado regime, streamwise vorticity clusters around $\sim$5 $\times$ 10$^{-3}$ s$^{-1}$, a magnitude associated with robust low-level mesocyclogenesis when sufficient stretching is present. In addition, the low LCLs, near-saturated inflow, and substantial low-level CAPE identified here provide favorable conditions for that amplification. Collectively, these results suggest that supercellular and non-supercellular tornadogenesis need not be mutually exclusive in the Philippine environment. Instead, both mechanisms may occur along a continuum, with locally generated, streamwise low-level horizontal vorticity, and its subsequent tilting and stretching providing a way for supercellular tornadogenesis and potential development of a persistent mesocyclone. The near-identical thermodynamic environments of tornadoes and waterspouts, distinguished primarily by subtle low-level kinematic differences and surface type, further support a continuum between these phenomena rather than strictly distinct dynamical regimes.

These findings carry a broader implication for how tropical severe-convective hazard is diagnosed globally. The tropics remain the region where reanalysis-based hazard reconstructions are least constrained. \citet{Battaglioli2026} explicitly caution that their globally applied very-large hail model, trained on three mid-latitude domains, should be interpreted carefully where regional climatologies cannot verify it or where ERA5 biases are present, singling out the tropics. Our results give that caution in the empirical form for the severe weather hazards being studied here. A tropical-maritime regime that is thermodynamically saturated and kinematically shallow will be systematically misclassified by parameters and models that weight DLS, including other kinematic parameters, and treat high CAPE as a discriminating rather than a background quantity. Thus, recalibration toward shallow-layer, orientation-aware kinematic discriminators such as 0--500 m to 0--1 km streamwiseness, streamwise vorticity, and helicity, with locally rescaled composite thresholds near SCP $\sim$ 1 and STP $\sim$ 0.5, recovers the diagnostic signal that the deep-layer formulation discards. This mirrors the trajectory of hail-environment research, where region-specific formulations such as the WMAXSHEAR discriminator with the BWD over the entire cloud depth \citep{Capuli2026} and China-calibrated STP and SCP for tornadoes \citep{RZhang2025} have materially outperformed their canonical parameters.

Several limitations bound these conclusions. The composite soundings derive from ERA5, whose representation of the convective boundary layer and of shallow near-surface shear is least certain in the tropics and over complex archipelagic terrain, precisely the layer on which our mechanistic argument rests. Thus, a direct evaluation against radiosonde or field-campaign profiles is a necessary next step. The event catalogue, though the most complete assembled for the Philippines, remains shaped by the reporting inhomogeneity common to all historical severe-weather databases, with the post-2019 surge attributable to heightened social attention and population density effect rather than a climatological signal; a pattern documented for Indonesia and elsewhere, so the climatology characterizes environments rather than absolute frequencies. Finally, the mean profiles smooth the substantial case-to-case variance visible in the underlying distributions and cannot by themselves establish the sufficiency of any single ingredient. Yet, the low-level streamwise vorticity mechanism we advance is consistent with the composites and with established theory but awaits confirmation from convection-permitting simulation and targeted observation. The suggested role of the archipelago's flanking mountain ranges in pre-conditioning the boundary layer for tornadogenesis likewise remains a spatial coincidence in need of dynamical testing. Notwithstanding these constraints, this study provides the first baseline climatology of tornadoes and waterspouts in the Philippines and, in doing so, adds a well-resolved tropical-maritime end-member to the global severe storm climatology.

\acknowledgments

The author (serving as the community project leader of Project SWAP) expresses sincere gratitude to the Philippine public for reporting the occurrence of tornadoes and waterspouts within the affected communities.
G. H. Capuli also acknowledges the constructive comments provided by the anonymous reviewers and the editor, which significantly improved the quality of this manuscript.
This work will serve as the last part of the community project; therefore, the additional data being added alongside DR4 will be presented in the Final Data Release (FDR). This work, including the community project, received no funding, but was `funded' by extensive and exhaustive effort, whose dedication and commitment to advancing our understanding of severe weather phenomena were indispensable. I am thankful to my loved ones for their unwavering support throughout this research.

\contribution

\textbf{Generich H. Capuli:} Writing - original draft, Writing - review \& editing, Visualization, Methodology, Investigation, Formal Analysis, Data curation, Conceptualization, Supervision.

\datastatement

The tornado and waterspout reports used in this study are archived through Project SWAP and its \href{https://doi.org/10.5281/zenodo.19178931}{DR4}; supplementary historical records added for this study will be released in the Final Data Release (FDR). ERA5 reanalysis data are openly available through the Copernicus \href{https://cds.climate.copernicus.eu/}{Climate Data Store}. The thunder-hour climatology was derived from Earth Networks Global Lightning Network (ENGLN) observations following \citet{digangi2022}; ENGLN data are commercial products of Earth Networks, and the derived thunder-hour fields can be obtained from the corresponding author upon reasonable request. The Philippine administrative boundaries are available from the Humanitarian Data Exchange (\href{https://data.humdata.org/dataset/cod-ab-phl}{HDX}). Proper attribution is required for these datasets.
\\
\\
This paper has made use of the following Python packages: \verb|ecape_parcel|, \verb|GeoPandas|, \verb|Matplotlib|, \verb|MetPy|, \verb|NumPy|, \verb|Pandas|, \verb|rioxarray|, \verb|SciPy|, \verb|seaborn|, and \verb|SounderPy|.

\interest

The author declares no competing interests.


\bibliographystyle{ametsocV6}
\bibliography{references}

\end{document}